\documentclass[prl,twocolumn,showpacs,amsmath,amssymb,letterpaper,floatfix]{revtex4}

\usepackage{graphicx}
\usepackage{dcolumn}
\usepackage{bm}

\begin{document}

\begin{minipage}[t]{4cm}
\includegraphics[width=3cm]{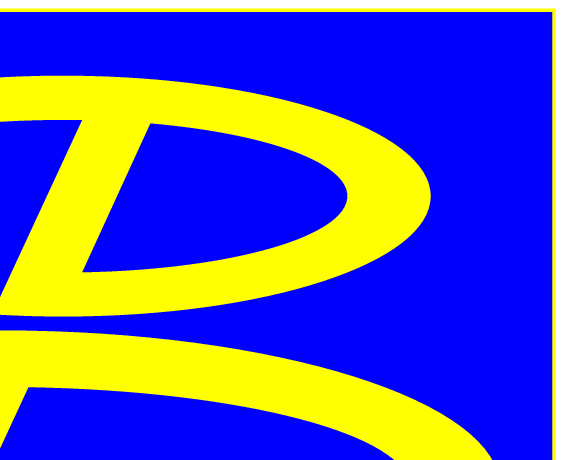}
\end{minipage}
\begin{minipage}[t]{10cm}
\hspace*{4cm}{KEK Preprint 2001-171}\\
\hspace*{4cm}{Belle Preprint 2002-5}
\end{minipage}

\title{
Observation of $\chi_{c2}$ Production in $B$-meson Decay
}

\author{Belle Collaboration}\noaffiliation


\author{
  K.~Abe$^{9}$,               
  K.~Abe$^{41}$,              
  T.~Abe$^{42}$,              
  I.~Adachi$^{9}$,            
  H.~Aihara$^{43}$,           
  M.~Akatsu$^{24}$,           
  Y.~Asano$^{48}$,            
  T.~Aso$^{47}$,              
  V.~Aulchenko$^{2}$,         
  T.~Aushev$^{14}$
% ...
}
\affiliation{
% ...
$^{2}${Budker Institute of Nuclear Physics, Novosibirsk}\\
% ...
$^{9}${High Energy Accelerator Research Organization (KEK), Tsukuba}\\
% ...
$^{14}${Institute for Theoretical and Experimental Physics, Moscow}\\
% ...
$^{24}${Nagoya University, Nagoya}\\
% ...
$^{41}${Tohoku Gakuin University, Tagajo}\\
$^{42}${Tohoku University, Sendai}\\
$^{43}${University of Tokyo, Tokyo}\\
% ...
$^{47}${Toyama National College of Maritime Technology, Toyama}\\
$^{48}${University of Tsukuba, Tsukuba}\\
% ...
}

\date{February 12, 2002}

\begin{abstract}

We report the first observation of $\chi_{c2}$ production in $B$-meson
decays. We find an inclusive $B\rightarrow \chi_{c2}X$ branching
fraction of $(1.80^{+0.23}_{-0.28}\pm 0.26)\times 10^{-3}$. The data set,
collected with the Belle detector at the KEKB $e^+e^-$ collider,
consists of 31.9 million $B\bar B$ events.  We also present branching
fractions and momentum spectra for both $\chi_{c1}$ and $\chi_{c2}$
production.

\end{abstract}

\pacs{13.25.Hw,14.40.Gx,14.40.Nd}

\maketitle

Although the theory for weak decays of $b$ quarks is formulated in
terms of quark processes, experiments are done with $B$ hadrons.  The
application of quantities calculated at the quark level to the
physically realizable hadrons usually requires theoretical assumptions
and approximations.  One widely used approximation is
``factorization,'' where it is assumed that the participating quarks
form hadrons with no subsequent transfer of quantum numbers between
them~\cite{FACT}.  Since this assumption is widely used, it is important that
the range of its validity is carefully tested.

In the factorization limit, decays of the type $B\rightarrow \chi_{c0}
X$ and $\chi_{c2}X$ are not allowed by angular momentum and
vector-current conservation~\cite{KNR}.  These decays can occur if
there is a (factorization-violating) exchange of soft gluons between
the quark pairs prior to hadron formation.  Belle has recently
reported the observation of the decay $B^-\rightarrow\chi_{c0} K^-$
with a decay branching fraction that is comparable to that for the
factorization-allowed decay $B^-\rightarrow J/\psi K^-$~\cite{CHIC0}.
The CLEO collaboration has published a 95\% CL upper limit on the inclusive
decay $B\rightarrow\chi_{c2}X$ of $2.0\times 10^{-3}$~\cite{CLEO}.

In this paper we report evidence for the inclusive decay $B\rightarrow
\chi_{c2} X$ from an analysis of 31.9 million $B\bar B$ events produced
in a 29.4 fb$^{-1}$ data sample taken at the $\Upsilon(4S)$ resonance
with the Belle detector at the KEKB asymmetric $e^+e^-$ collider.  An
additional 3.0 fb$^{-1}$ sample taken at a center-of-mass
energy 60~MeV below the $\Upsilon(4S)$ is used to study backgrounds
from non-resonant (continuum) processes.

The Belle detector consists of a
three-layer silicon vertex detector (SVD), a 50-layer central drift
chamber (CDC), an array of aerogel threshold $\rm\check{C}erenkov$
counters (ACC), time-of-flight scintillation counters (TOF), a CsI(Tl)
crystal electromagnetic calorimeter (ECL), a 1.5 T superconducting
solenoid coil and an instrumented iron-flux return for muon and $K_L$
detection (KLM).  The detector is described in detail
elsewhere~\cite{Belle}.

Events with candidate $B$ mesons are selected by first applying
general hadronic event criteria. These include the requirement of at
least three charged tracks, an event vertex consistent with the
interaction point, reconstructed center-of-mass (CM) energy greater
than $0.2\sqrt{s}$, a longitudinal component of reconstructed CM momentum
less than $0.5\sqrt{s}/c$, and a total ECL energy between $0.1\sqrt{s}$
and $0.8\sqrt{s}$ with at least two energy clusters.
To suppress continuum backgrounds
we also require the ratio of the second to zeroth Fox-Wolfram
moments to be less than 0.5~\cite{FoxWolf}.

We reconstruct $\chi_{c1}$ and $\chi_{c2}$ via the decays to
$J/\psi\gamma$, {$J/\psi\rightarrow l^+l^-$}. Both leptons are required
to be loosely identified as leptons. Electrons are identified using a
combination of drift chamber $dE/dx$ measurements, aerogel response, and
electromagnetic shower position, shape and energy. Muons are
identified with KLM hit positions and penetration depth. In order to
recover dielectron events where one or both electrons have radiated a
photon (final state radiation or bremsstrahlung),
we include the four-momentum of every photon detected within 0.05
radians of the original $e^+$ or $e^-$ direction in the invariant mass
calculation.
The
$J/\psi\rightarrow\mu^+\mu^-(e^+e^-)$ candidate invariant mass is
required to be between $-25(-40)$ MeV/$c^2$ and $+25$ MeV/$c^2$ of the known
$J/\psi$ mass, with an expected resolution of 9.6(10.8) MeV/$c^2$ for dimuon(dielectron)
$J/\psi$'s.  The larger range for dielectron candidates
is to include candidates that fall in the radiative tail, even after
the photon correction.

To reduce combinatoric background, we veto gamma candidates that form a
good $\pi^0$ candidate with any other photon candidate of energy
greater than 60 MeV in the event. A good candidate $\pi^0$ is defined by a
$\chi^2$ of less than 10 after a mass-constrained kinematic fit.
We then make a histogram of the mass difference between the $\chi_c$
and the $J/\psi$ candidates; this nearly eliminates the effect of the
$J/\psi$ measurement error.
The error on the mass difference is dominated by the photon energy
resolution.
The momentum of
the $\chi_c$ candidate in the CM reference frame is required to be
less than 1.7 GeV/$c$ (the kinematic limit for a $\chi_c$ coming from a
$B$ meson); this requirement was not used in the determination of
the $\chi_c$ momentum spectra.

\begin{figure}
\includegraphics[width=0.9\linewidth]{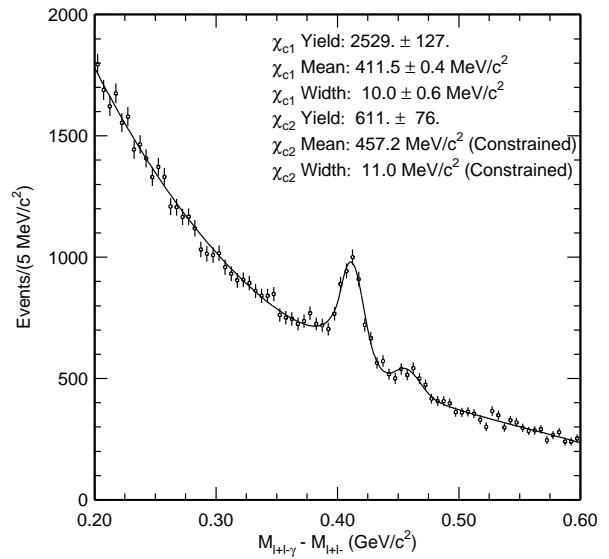}
\caption{\label{fig:yield}
The $\chi_c-J/\psi$ mass difference distribution for candidate events.
The widths reported correspond to the ``width'' parameter of the
Crystal Ball function~\cite{CB}.
}
\end{figure}

In Fig.~\ref{fig:yield} a clear $\chi_{c2}$ peak can be seen next to a
larger $\chi_{c1}$ peak. In order to determine the yield we fit the
distribution to two Crystal Ball line shapes~\cite{CB}
and a third-order Chebyshev polynomial for the background.
The Crystal Ball function allows for a ``tail'' in the line shape that
is due to photon shower leakage in the ECL.

In this fit (the ``standard'' fit), the signal line shapes (i.e. the
widths, means, and tail parameters) are allowed to float with the following
constraints: the difference between the means is fixed to the known
$\chi_{c1}-\chi_{c2}$ mass difference; the $\chi_{c2}$ width is fixed
to 1.1 times the $\chi_{c1}$ width, to take into account the Monte Carlo
expected ratio of the widths, which is consistent with a higher
average $\chi_{c2}$ photon energy; and the tail parameters are fixed to
be the same. The background shape is fixed by fitting to the regions
outside the signal region from 0.35 to 0.50 GeV/$c^2$.

The signal shape was compared with predictions from an inclusive
$B\rightarrow \chi_{c1}X$ and $\chi_{c2}X$ full Monte Carlo
simulation.  The signal widths in data are larger.
In a study of $D^{*0}\rightarrow D^0\gamma$, we find that the
calorimeter response for a single photon is broader in the data than
in the Monte Carlo: for the $\chi_c - J/\psi$ mass difference, we
expect the width to be increased by a factor of 1.3.
For
$\chi_{c1}$ the Monte Carlo width is $7.0\pm 0.2$ MeV/$c^2$, the
corrected width is $9.1\pm 0.3$ MeV/$c^2$, and the measured width is
$10.0\pm 0.6$ MeV/$c^2$.
We consider the variation in signal
yields for various fitting scenarios in determining the systematic
error due to fitting.

The background shape was checked against a full Monte Carlo simulation
that included the appropriate amounts of $B\bar B$ and non-resonant
events. The Monte Carlo and data background shapes are in
good agreement and their normalizations agree within 3\%.

We find a yield of $2529\pm127$ events in the $\chi_{c1}$ peak and
$611\pm76$ events in the $\chi_{c2}$ peak, where the error is
statistical only.

Several sources of background production were checked.
Two-photon processes produce $\chi_{c2}$~\cite{2Photon}.
To estimate the contribution to the $\chi_{c2}$ signal from
events of this type we looked
at the equivalent of 560 fb$^{-1}$ of Monte Carlo data. From this 
sample we estimate a background contribution of 1.9 events.
We also checked the
3.0 fb$^{-1}$ continuum data sample for $\chi_c$ production. We expect
a small number of events from feed down from continuum $\psi(2S)$'s
and possible direct $\chi_{c}$ production.
From the fit we find $14.0\pm 6.4$ events in the $\chi_{c1}$ region and $0.4\pm
5.7$ events in the $\chi_{c2}$ region.
Expected contributions of feed down from continuum $\psi(2S)$
production are 0.5 events for $\chi_{c1}$ and 0.2 events for $\chi_{c2}$~\cite{CONT},
and, hence, consistent with the above measurements.
For the $\chi_{c2}$ case, we
follow the prescription of Feldman and Cousins and find the
68.27\% confidence interval for the event yield to be [0.0,6.1]~\cite{FC}. 
We scale the continuum yields by the ratio of on- and off-resonant
luminosities, corrected for the difference in continuum cross section
due to the slight difference in beam energies. The scaled $\chi_{c1}$
and $\chi_{c2}$ continuum yields are subtracted from the on-resonance
yields. We use the Feldman-Cousins confidence limits in determining
the statistical error for $\chi_{c2}$ after the subtraction.

To convert yields to branching fractions we determine
the reconstruction efficiency with a full inclusive
$B\rightarrow\chi_{c1}X$ and $\chi_{c2}X$ Monte Carlo.
We find the efficiencies for reconstruction to be
$32.0\pm 0.5$\% and
$33.1\pm 0.9$\%, respectively.
The $\chi_c$ momentum spectra of the Monte Carlo are similar
to those measured in data. The efficiencies are
uniform over the allowed $\chi_{c1}$, $\chi_{c2}$ momentum range.

We use the 2001 Particle Data Group~\cite{PDG} values for
daughter branching fractions
${\cal B}$$(J/\psi\rightarrow l^+l^-) = 0.118\pm 0.002$,
${\cal B}$$(\chi_{c1}\rightarrow J/\psi\gamma) = 0.273\pm 0.016$, and
${\cal B}$$(\chi_{c2}\rightarrow J/\psi\gamma) = 0.135\pm 0.011$.
The inclusive $B\rightarrow
\chi_{c}X$ branching fractions are found to be:
${\cal B}$$(B\rightarrow \chi_{c1}X) = (3.63\pm 0.22)\times 10^{-3}$, and
${\cal B}$$(B\rightarrow \chi_{c2}X) = (1.80^{+0.23}_{-0.28})\times 10^{-3}$.
These numbers are summarized in Table~\ref{tab:yields}.

\begin{table*}
\caption{\label{tab:yields}Yields and branching fractions.
Errors are statistical only.}
\begin{ruledtabular}
\begin{tabular}{lcccc}
                     & \multicolumn{2}{c}{$\chi_{c1}$}& \multicolumn{2}{c}{$\chi_{c2}$} \\
                     & Yield         & BF ($10^{-3}$) & Yield       & BF ($10^{-3}$) \\
\colrule					      				    
Fit                  & $2529\pm 127$ & ---            & $611\pm 76$ & ---            \\
Continuum subtracted & $2391\pm 142$ & $3.63\pm 0.22$ & $607^{+76}_{-94}$ & $1.80^{+0.23}_{-0.28}$ \\
Feed down subtracted & ---           & $3.32\pm 0.22$ & ---               & $1.53^{+0.23}_{-0.28}$ \\
\end{tabular}
\end{ruledtabular}
\end{table*}

Some of the $B\rightarrow \chi_c$ decays result from ``feed down''
from the $\psi(2S)$; these are not forbidden by factorization. In
order to determine the rate for direct decays to the $\chi_c$ states,
the $\psi(2S)$ contribution must be subtracted.
This feed down is estimated using the Particle Data Group $B\rightarrow \psi(2S) X$
and $\psi(2S)\rightarrow \chi_c \gamma$ branching fractions.
After correcting for feed down we find:
${\cal B}$$(B\rightarrow \chi_{c1}X) = (3.32\pm 0.22)\times 10^{-3}$, and
${\cal B}$$(B\rightarrow \chi_{c2}X) = (1.53^{+0.23}_{-0.28})\times 10^{-3}$.

Significant sources of systematic error are in the
efficiencies for lepton identification (2\% per lepton track),
tracking (2\% per track), photon detection (2\%), as well as
daughter branching fractions (6\% for $\chi_{c1}$, 8\% for $\chi_{c2}$),
and fitting systematics (4\% for  $\chi_{c1}$, 10\% for $\chi_{c2}$).
The systematic errors are summarized in Table~\ref{tab:syserr}.

\begin{table}
\caption{\label{tab:syserr}Systematic Errors.}
\begin{ruledtabular}
\begin{tabular}{lcc}
                      & $\chi_{c1}$  & $\chi_{c2}$ \\
\colrule
Lepton identification &  4\%         & 4\%         \\
Tracking efficiency   &  4\%         & 4\%         \\
Photon efficiency     &  2\%         & 2\%         \\
${\cal B}(\chi_c)$    &  6\%         & 8\%         \\
Monte Carlo Statistics &  1\%         &  3\%        \\
Fit                   &  4\%         & 10\%        \\
\colrule
Total                 &  9\%         & 14\%        \\
\end{tabular}
\end{ruledtabular}
\end{table}

The fit for the $\chi_{c1}$ and $\chi_{c2}$ yields is sensitive to the
signal and background shapes. We estimate the error associated with
the fit by performing the fit in a variety of ways including: fixing
the signal means, widths, and tail shapes to Monte Carlo values (with
the widths multiplied by a scaling factor and separately by
adding a random number from a Gaussian distribution generated
to yield the desired width increase); allowing the means to float,
with the widths and tail shape fixed; allowing the means and widths to
float, with the tail shape fixed; and allowing all parameters to
float. In all cases, when a parameter is allowed to float, the
$\chi_{c1}$ and $\chi_{c2}$ line shapes are constrained appropriately
as with the standard fit. Two methods of fitting the backgrounds are
also used: fixing the background with the sidebands (as with the
standard fit) and allowing the background shape to float freely. The
one combination that is not used is to fit with a free tail shape
and a free background shape as there can be a trade off between the
background area and tail area in the fit.

In addition to the above fits, we confirmed that a third-order
polynomial is sufficient to fit the background by performing a fit to
the background Monte Carlo; adding additional terms did not improve
the confidence levels of the fits. The fitting systematic error is
assigned from the largest variation between the fits described above and our
standard fit.

\begin{figure}
\includegraphics[width=0.9\linewidth]{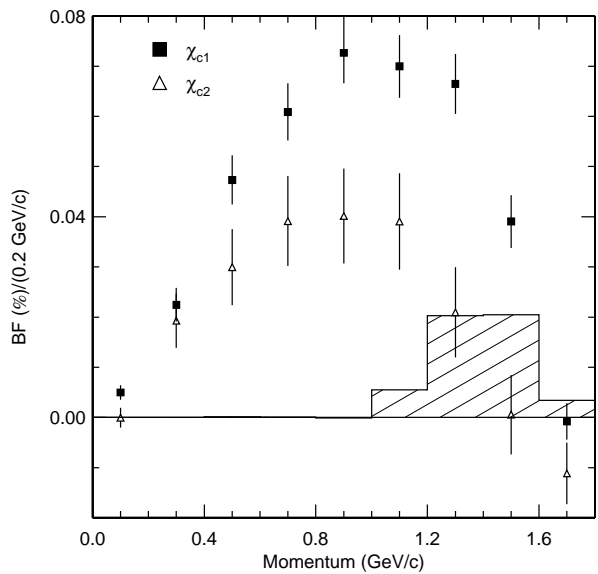}
\caption{\label{fig:p_bf}
Branching fractions for
$B\rightarrow \chi_{c1} X$ and
$B\rightarrow \chi_{c2} X$ as a function of $\chi_c$ momentum
in the $e^+e^-$ center-of-mass frame.
Background from
continuum processes and feed down from $\psi(2S)$ have not been subtracted.
The shaded region has the expected shape for a contribution from $B\rightarrow \chi_{c2}K$.}
\end{figure}

The $\chi_c$ momentum spectra are interesting as they can give clues
to the production mechanisms. The high momentum end is dominated by two-body
decays to $\chi_{c1}(\chi_{c2}) K$ and $\chi_{c1}(\chi_{c2}) K^*$
while the low end may be from higher mass $K^*$
resonances, multi-body decays or feed down from $\psi(2S)$. To determine
the momentum spectra, we divide the data into sets based on the
momentum of the $\chi_c$ candidate.  We then fit each distribution for
the $\chi_{c1}$ and $\chi_{c2}$ yields,
which are converted into
differential branching fractions, corrected bin-by-bin for the detector
efficiency. The resulting momentum spectra, shown in
Fig.~\ref{fig:p_bf}, are broad indicating that a large
component of either multi-body decays or higher $K^*$ resonances
is present.
The shaded histogram in Fig.~\ref{fig:p_bf} shows the $\chi_{c2}$
momentum distribution for Monte Carlo-simulated $B\rightarrow\chi_{c2}K$
decays, which indicates that almost all $\chi_{c2}$'s from
these decays have momenta between 1.2 and 1.6 GeV/$c$.
After doing a fit of this Monte
Carlo histogram to the data histogram we find an upper limit at
the 90\% confidence level of $5.0\times 10^{-4}$ for the
$B\rightarrow \chi_{c2}K$ branching fraction.
The shaded area in Fig.~\ref{fig:p_bf} corresponds to this upper limit.
A more detailed analysis of this decay is forthcoming.

In summary,
we report the first statistically significant observation of
$\chi_{c2}$ production in $B$-meson decays. The
$B\rightarrow \chi_{c1} X$ and $B\rightarrow \chi_{c2}X$ branching
fractions are measured to be
$(3.63        \pm 0.22\pm 0.34)\times 10^{-3}$ and
$(1.80^{+0.23}_{-0.28}\pm 0.26)\times 10^{-3}$, respectively, where the
first error is statistical and the second systematic.
After subtraction for feed down from $\psi(2S)$, we find the direct
branching fractions to be
$(3.32        \pm 0.22\pm 0.34)\times 10^{-3}$ and
$(1.53^{+0.23}_{-0.28}\pm 0.27)\times 10^{-3}$ respectively.
The non-zero
$\chi_{c2}$ production is an indication that the factorization model
does not give a complete picture for charmonium production in
$B$-meson decays. The momentum spectra include a large low momentum
component, indicating either multibody final states or final states
with higher resonant $K^*$ production.

We wish to thank the KEKB accelerator group for the excellent
operation of the KEKB accelerator.
We acknowledge support from the Ministry of Education,
Culture, Sports, Science, and Technology of Japan
and the Japan Society for the Promotion of Science;
the Australian Research Council
and the Australian Department of Industry, Science and Resources;
the National Science Foundation of China under contract No.~10175071;
the Department of Science and Technology of India;
the BK21 program of the Ministry of Education of Korea
and the CHEP SRC program of the Korea Science and Engineering
Foundation;
the Polish State Committee for Scientific Research
under contract No.~2P03B 17017;
the Ministry of Science and Technology of the Russian Federation;
the Ministry of Education, Science and Sport of Slovenia;
the National Science Council and the Ministry of Education of Taiwan;
and the U.S.\ Department of Energy.

\end{document}